\documentclass[a4paper]{ESASPCS13Style}
\usepackage{epsfig}

\def\vsini{$v$\,$\sin i$}

\def\rsol{R$_{\odot}$}
\def\micron{$\mu$m}
\def\teff{${\rm T}_{\rm eff}$}

\begin{document}

\title{Variability and rotation of ultra cool dwarfs}

\author{C.A.L.\ Bailer-Jones} \institute{Max-Planck-Institut f\"ur Astronomie, K\"onigstuhl 17, 69117
Heidelberg, Germany}

\maketitle 

\begin{abstract}
Over the past few years monitoring programs have shown ultra cool
dwarfs (UCDs) to be photometrically variable. Of the 60 sources
monitored in the field and some 120 monitored in clusters, about 40\% show
variability in both cases.  For mid to late M dwarfs in young ($<$100
Myr) clusters, this variability is generally periodic with amplitudes
of up to a few tenths of a magnitude and periods of between a few hours and
several days.  For older field dwarfs (covering late M, L and T types)
this variability is often nonperiodic with smaller amplitudes (up to
0.1 mag in I) and timescales of order hours. The former may be
attributed to the rotational modulation of magnetically-induced
photospheric spots, as seen in higher mass T Tauri stars. The
nonperiodic variability, on the other hand, may be caused by a rapid
evolution of surface features (which `mask' the otherwise observable
rotational modulation). This could be related to the formation and
dissipation of inhomogeneities in dust clouds in the photospheres of
UCDs.

\end{abstract}

Work over the past few years by several groups has shown strong
evidence for low amplitude photometric variability in both field and
cluster very low mass stars and brown dwarfs (collectively, {\it ultra
cool dwarfs}, or UCDs).  An example light curve and power spectrum is
shown in Fig.\ 1.

\begin{figure}[t]
\begin{center}
\epsfig{figure=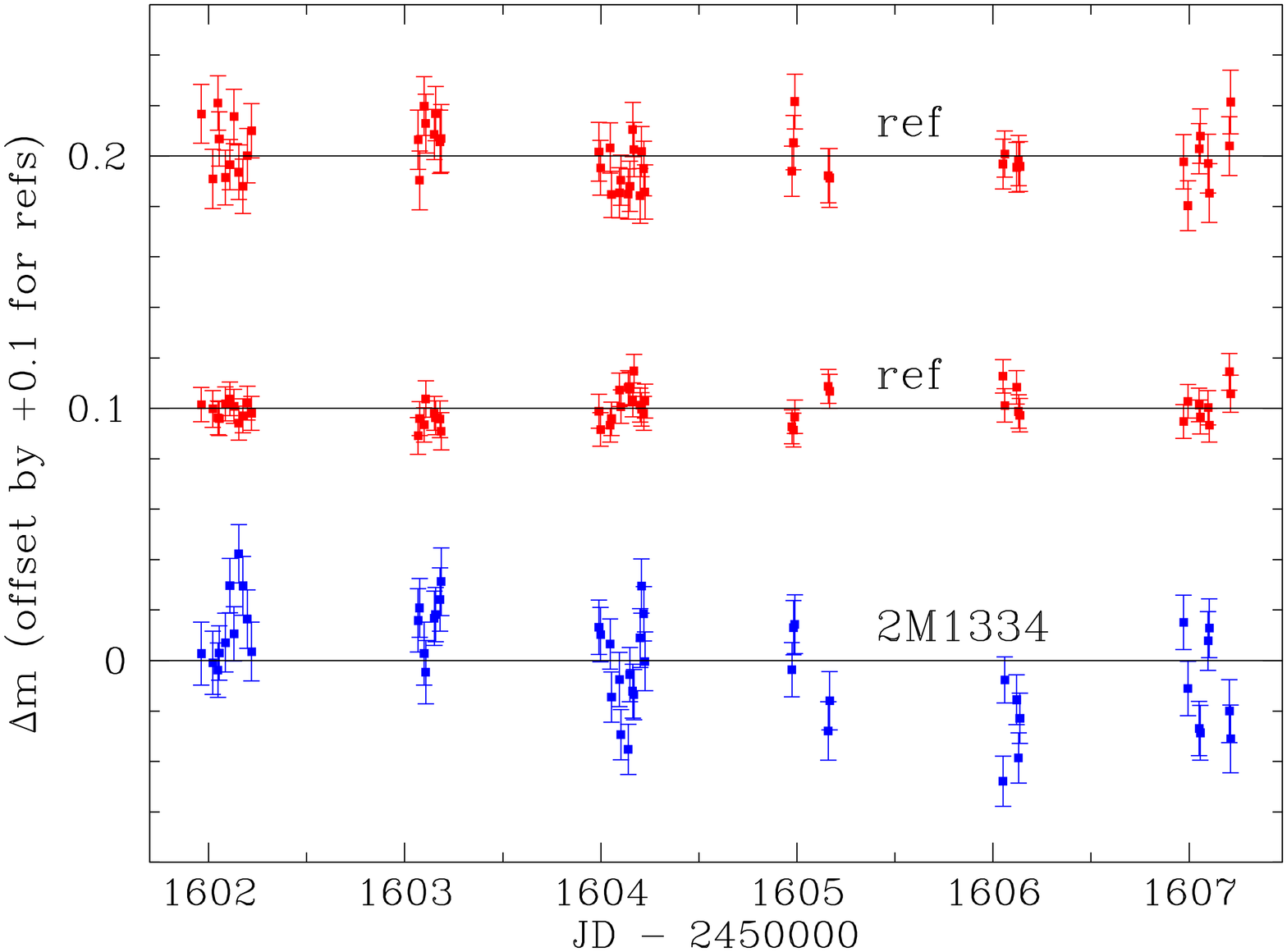, width=6cm, angle=-90}
\vspace*{2ex}
\epsfig{figure=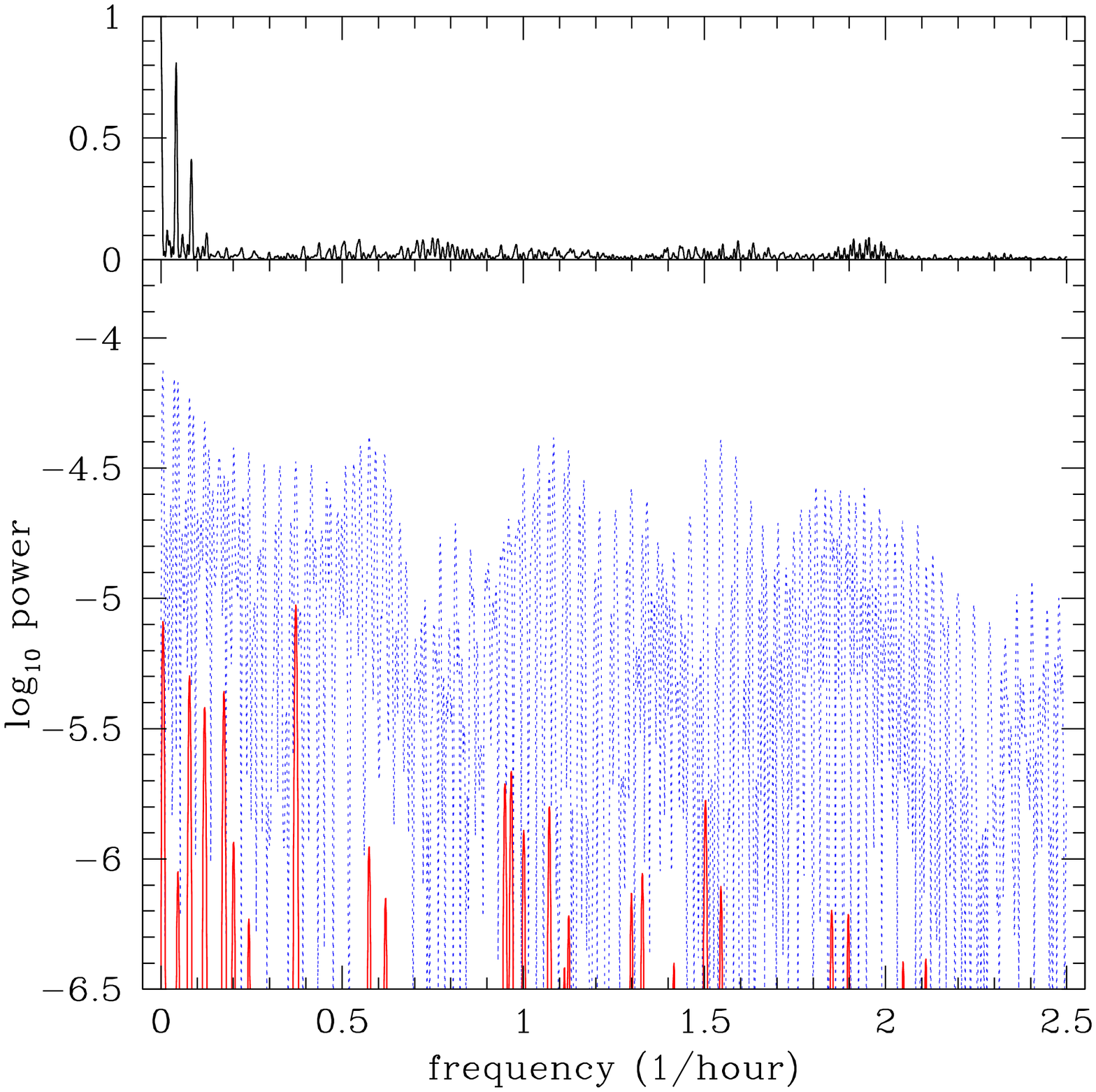, width=6cm, angle=0}
\end{center}
\caption{The top panel shows the differential light curve of a field
L dwarf (bottom) and two (of many) reference stars (top two). The
bottom panel shows the corresponding periodogram for the L dwarf, with
the raw, or dirty periodogram in blue (dashed) and the CLEANed
periodogram in red (solid). The window function is shown above.}
\label{fig1}
\end{figure}

For the field UCDs, variability timescales are typically of order a
few hours with amplitudes of between 0.01 and 0.08 mags in the I band.
Fig.\ 2 shows the variability detection amplitudes and upper limits as
a function of spectral type. There is no particular correlation, also
not with the amplitudes against spectral type.
 
\begin{figure}[t]
\begin{center}
\epsfig{file=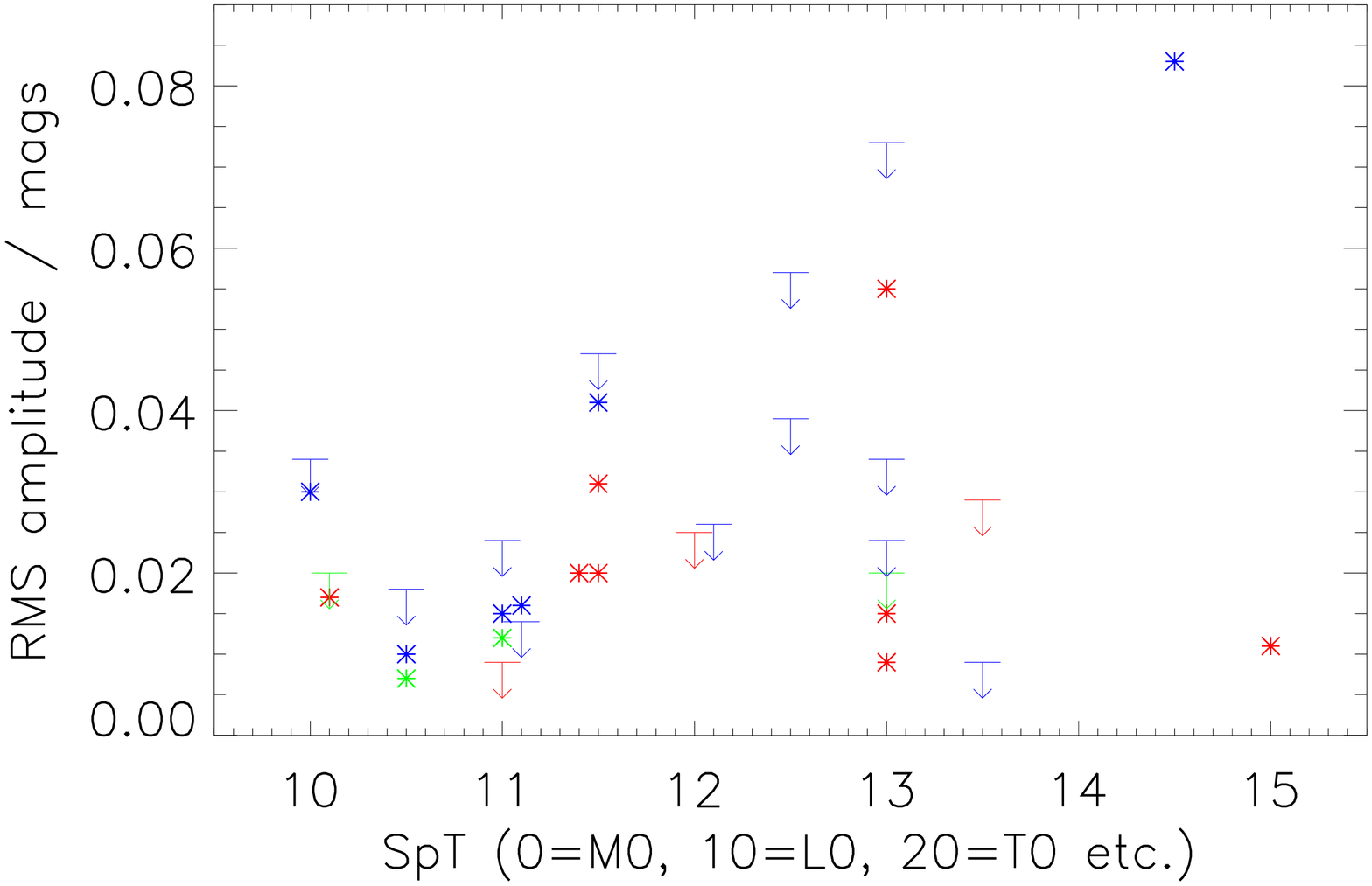, width=6cm, angle=90}
\epsfig{file=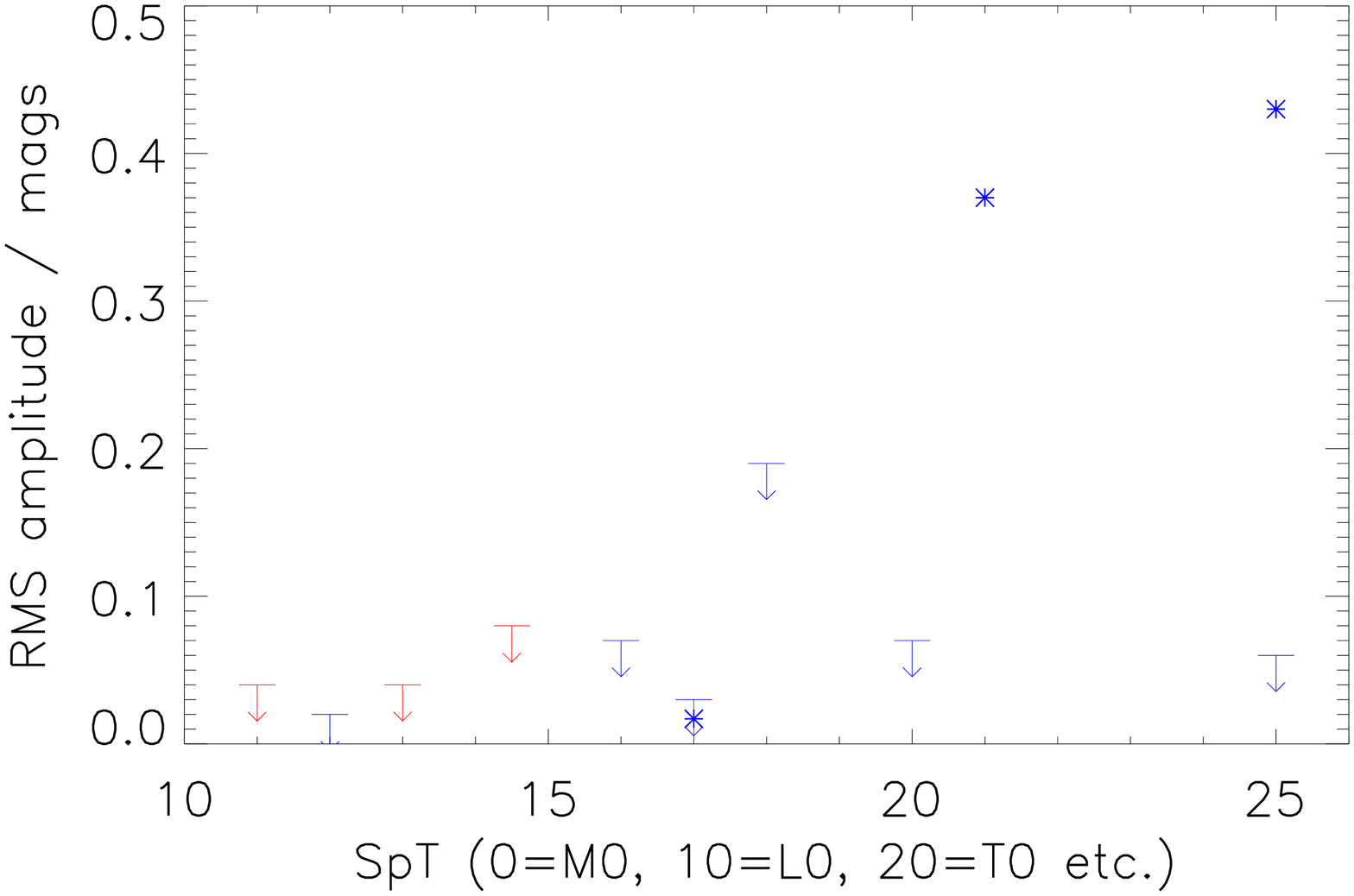, width=6cm, angle=90}
\end{center}
\caption{Variability detection amplitudes (stars) and upper limits
(arrows) as a function of spectral type and magnitude for field
UCDs. The data are taken from five works in the literature:
Bailer-Jones \& Mundt (2001) [red], Clarke et al.\ (2002) [green],
Gelino et al.\ (2002) [blue] (I-band surveys, upper panel);
Bailer-Jones \& Lamm (2003) [red] and Enoch et al.\ (2002) (K-band
surveys, lower panel).}
\label{fig2}
\end{figure}

In several cases, the variability in UCDs has been found to be
non-periodic.  This is curious, as in many cases the monitoring
surveys would have been sensitive to expected UCD rotation periods
(see Fig.\ 3).  Bailer-Jones \& Mundt (2001) interpreted this with a
{\it masking hypothesis}: If surface evolve on a timescale shorter
than the rotation period, these will obscure a regular modulation of
the light curve.

\begin{figure}[t]
\begin{center}
\epsfig{file=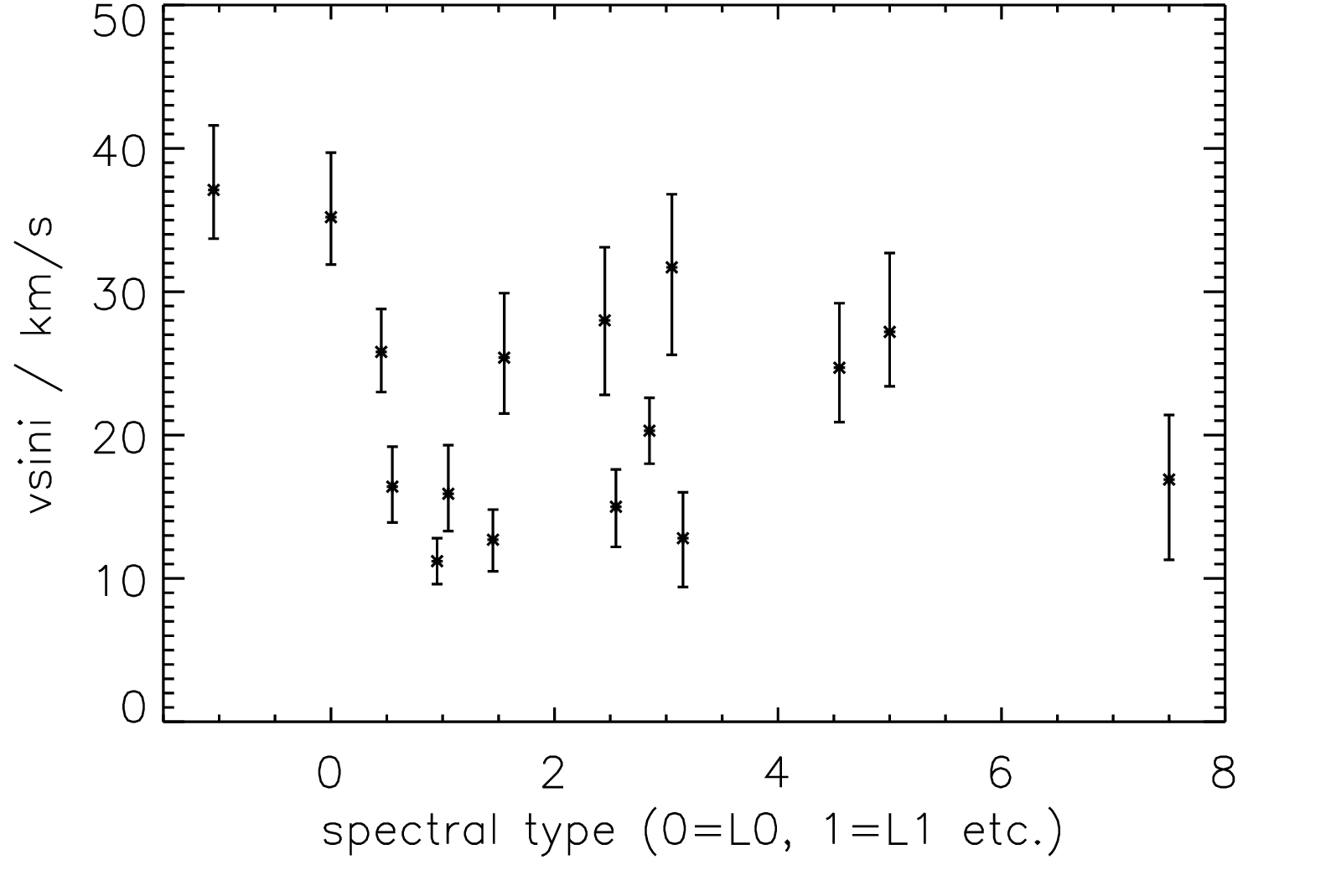, width=5cm, angle=90}
\vspace*{2ex}
\epsfig{file=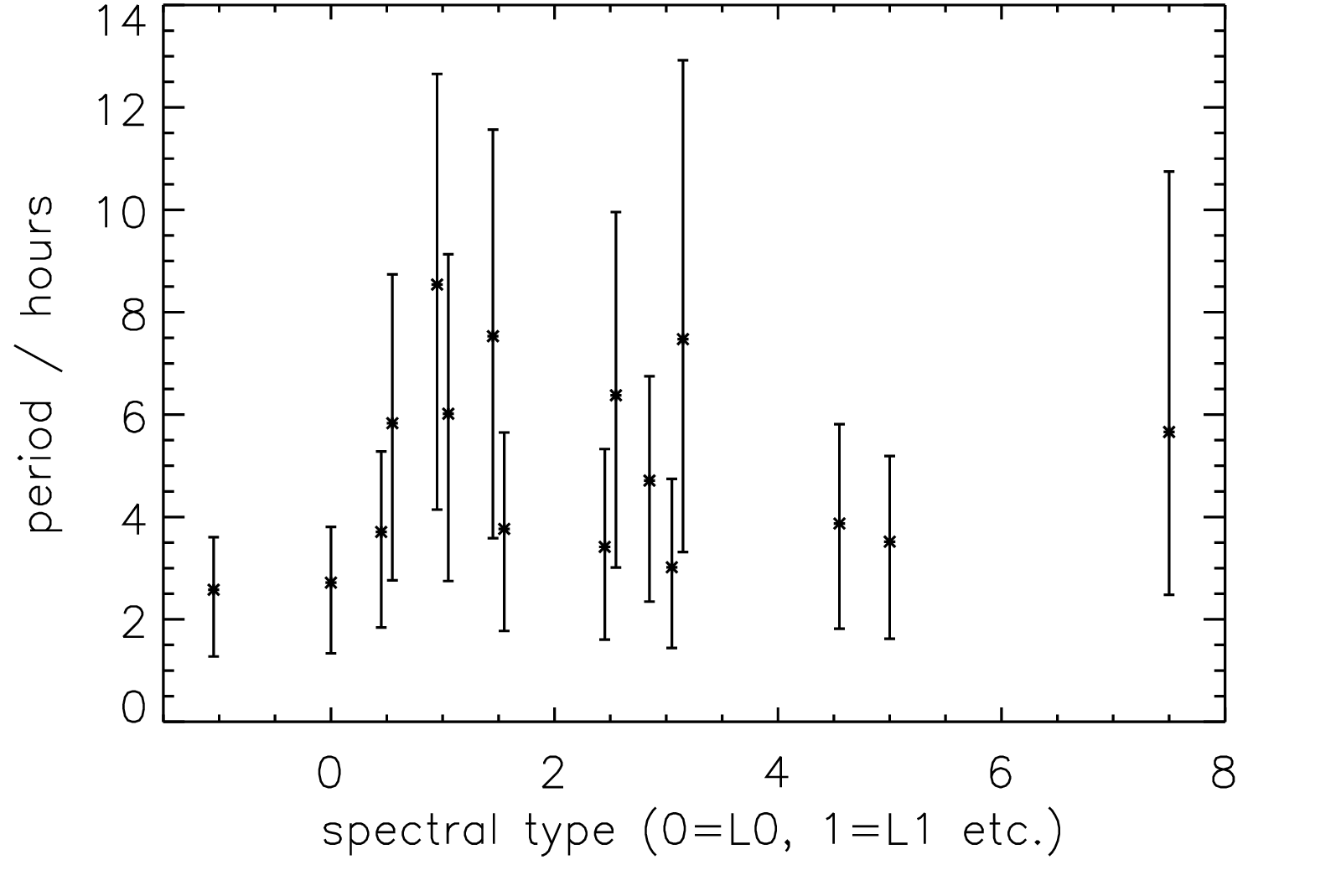, width=5cm, angle=90}
\end{center}
\caption{Bailer-Jones (2004) determined the projected rotation
velocities (\vsini) of 16 field UCDs using VLT/UVES. The top panel
shows these to be rapid rotators (the error bars show the {\it
maximum} uncertainties, not the 1$\sigma$ ones). Adopting a radius of
0.1\,\rsol\ for UCDs, these are converted into expected periods,
maximum rotation periods and 90\%-confidence limits for the minimum
periods, plotted in the bottom panel. From this, claimed photometric
periods in the literature could be confirmed or refuted. In one case
these data permit a determination of the lower limit of a UCD radius
as $0.097 \pm 0.007$\,\rsol.}
\label{fig3}
\end{figure}

An analysis of the literature shows that UCD variability is quite
common.  I have collated these results into Tables 1 and 2 for field
UCDs and cluster UCDs respectively. My literature search tries to
include all relevant work in the refereed literature as of mid
2004. All surveys are in the I-band, except Enoch et al.\ 2002 (K
band), Bailer-Jones \& Lamm 2003 (J and K bands) and Tinney \& Tolley
1999 (two narrow bands).  The issue of a detection or non-detection is
somewhat arbitrary for marginal cases, as it depends on the
statistical test and threshold used. As far as is possible, I have
converted values to a common 99\% confidence level for variability
detection. One must obviously be very careful in comparing results
from different surveys, as they differ in their sensitivity limits,
target selection, spectral types etc. Keeping this in mind, it appears
that both cluster and field UCDs show a similar variability fraction
of 30--40\% down to optical and near infrared amplitudes greater than
0.5--1\%.  However, cluster UCDs are more likely to show periodic
variability -- presumably a rotational -- whereas the field UCDs often
display non-periodic variability.  Thus while both sets are rapid
rotators, it seems that surface features evolve more rapidly on field
UCDs.

\begin{table}[bht]
\caption{Variability detections and non-detections in field ultra cool
dwarfs, taken from the literature.  The three columns show: the number
of variables; the number of non-variable; the variability fraction (of
the total). These are summed at the bottom (the total percentage
is calculated from the other two totals).  Numbers may deviate slightly
from those published as I have attempted to adopt a common 99\%
confidence level for detecting variability.}
\label{tab1}
\begin{center}
\leavevmode
\footnotesize
\begin{tabular}[h]{lrrr}
\hline \\[-5pt]
Reference                  & \#var   & \#non  &   fraction\\[+5pt]
\hline \\[-5pt]
Bailer-Jones \& Mundt 2001 & 8 & 3  & 73\% \\
Bailer-Jones \& Lamm 2003  & 0 & 3  & 0\% \\
Clarke et al.\ 2002a       & 2 & 2  & 50\% \\
Clarke et al.\ 2002b       & 1 & 0  & 100\% \\
Enoch et al.\ 2002         & 3 & 6  & 33\% \\
Gelino et al.\ 2002        & 6 & 12 & 33\% \\
Koen 2003                  & 3 & 9  & 25\% \\
Martin et al. 2001         & 1 & 0  & 100\% \\
Tinney \& Tolley 1999      & 1 & 1  & 50\% \\[+5pt]
\hline \\[-5pt]
TOTAL                      & 25   & 36    &   41\% \\[+5pt]

\hline \\
\end{tabular}
\end{center}
\end{table}

\begin{table}[bht]
\caption{As Table 1, but for UCDs in clusters. The name of the cluster
is given in the second column: P=Pleiades, S=$\sigma$~Orionis,
C=Chamaeleon I. Because the number of non-detections in Scholz \&
Eisl\"offel (2004b) is not entirely clear, and because these figures
would dominate the sum, these results have been excluded from the
total (but are included in the totals shown in parentheses).}
\label{tab2}
\begin{center}
\leavevmode
\footnotesize
\begin{tabular}[h]{llrrr}
\hline \\[-5pt]
Reference                  &  & \#var   & \#non  &   fraction\\[+5pt]
\hline \\[-5pt]
Joergens et al.\ 2003        & C &  5 &   5 & 50\% \\
Bailer-Jones \& Mundt 2001   & P & 1  & 4  & 20\% \\
Scholz \& Eisl\"oeffel 2004a & P & 12 & 14 & 46\% \\
Terndrup et al.\ 1999        & P &  2 &  6 & 25\% \\
Bailer-Jones \& Mundt 2001   & S &  3 &  3 & 50\% \\
Caballero et al.\ 2004       & S & 11 &  17 & 39\% \\
Scholz \& Eisl\"oeffel 2004b & S & (23) &  (~72) & (24\%) \\
Zapatero Osorio et al. 2003  & S &  1 &  0 & 100\% \\[+5pt]
\hline \\[-5pt]
TOTAL                        &   & 35   & 49    &   42\% \\
                             &   & (58)   & (121)    &   (32\%) \\[+5pt]
\hline \\
\end{tabular}
\end{center}
\end{table}

There are at least two plausible candidates for causing the
variability.  The first is cool, magnetically-induced spots.  This is
an attractive explanation for cluster UCDs: as these are young, they
may show activity with spots appearing in
analogy to weak-lined T Tauri stars. However, Gelino et al.\ (2002)
and Mohanty et al.\ (2003) have argued against the presence of spots
at these very low temperatures because of the neutrality of the
photosphere and thus a weak coupling between the gas and any magnetic
field. A second explanation is dust clouds. Dust is known to form at
these low temperatures. Rapid rotation and convection could give rise
to complex atmospheric dynamics, possibly accounting for the non-periodic
variability seen in field L and T dwarfs.

\begin{figure}[t]
\begin{center}
\epsfig{file=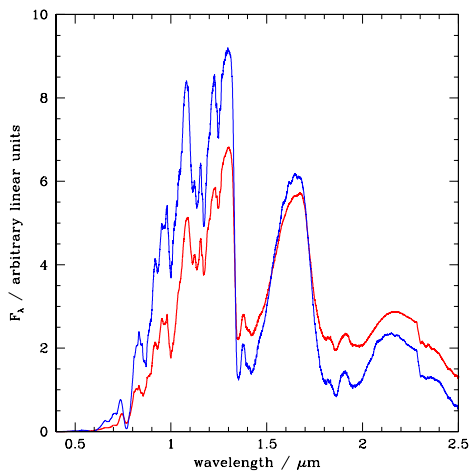, width=6.5cm, angle=0}
\end{center}
\caption{Model DUSTY (red/thick) and COND (blue/thin) spectra from Allard et
al.\ (2001) for a UCD with \teff\,=\,1900\,K.}
\label{fig4}
\end{figure}

\begin{figure}[t]
\begin{center}
\epsfig{file=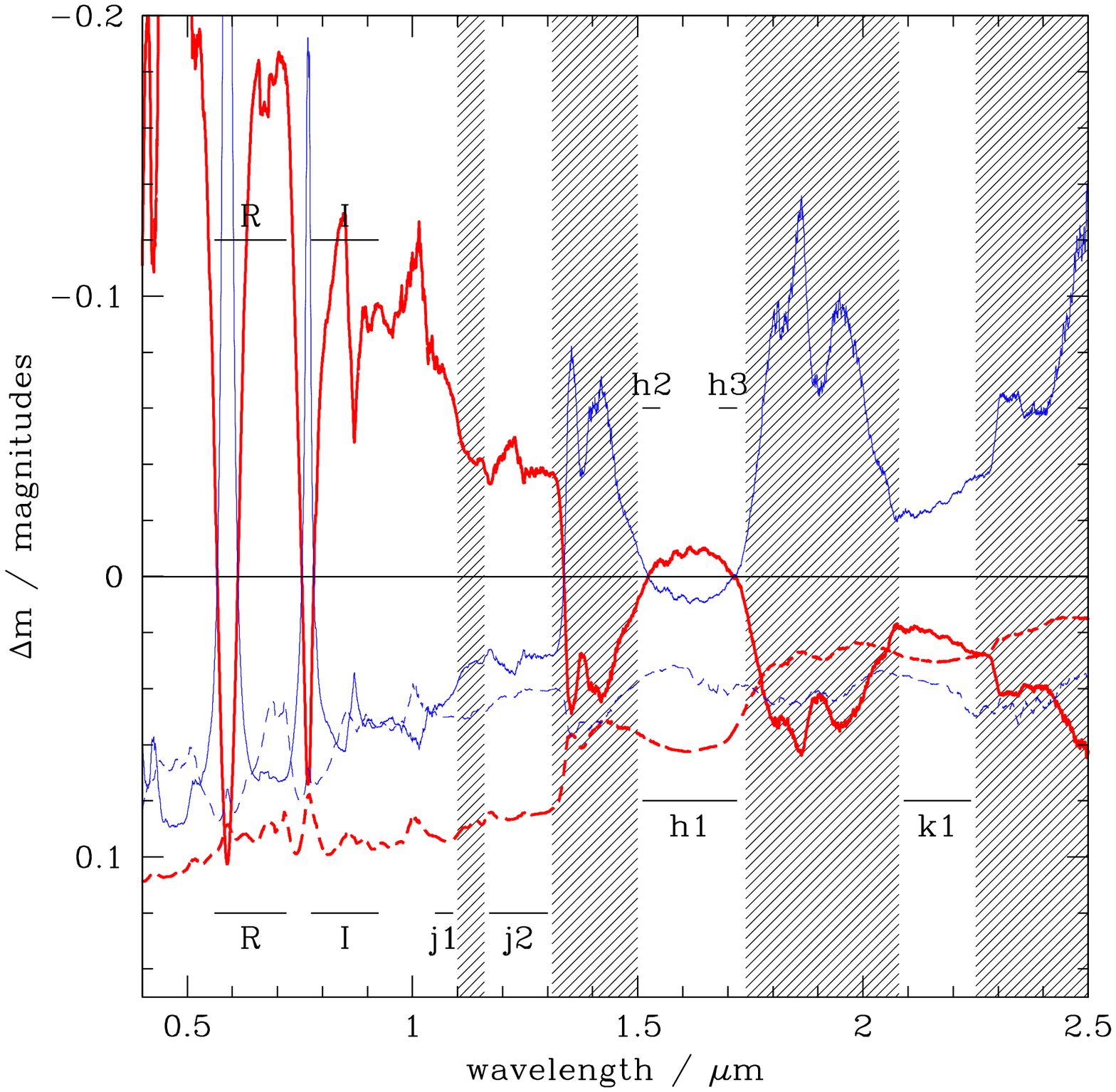, width=7.5cm, angle=0}
\end{center}
\caption{Predictions of the change in the spectrum of a UCD due to
the formation of a cloud or spot with a 10\% filling fraction
(Bailer-Jones 2002).  The four lines shown are for: COND cloud on a
DUSTY atmosphere (thick/red solid line) and 200\,K cooler spot on a
DUSTY atmosphere (thick/red dashed line); DUSTY cloud on a COND
atmosphere (thin/blue solid line) and 200\,K cooler spot on a COND
atmosphere (thin/blue dashed line).}
\label{fig5}
\end{figure}

I have made initial attempts to predict and observe the spectroscopic
signatures of different types of spot and cloud patterns (Bailer-Jones
2002), as shown in Figs.\ 4 and 5.  To test these predictions, I
obtained time-resolved differential spectrophotometric observations of
one field L1.5 dwarf. Spectra were obtained relative to a reference
star observed simultaneously in the slit (see Fig.\ 6). There is no
strong evidence for variability in any {\it single} band, but there is
evidence for colour-correlated variability (Fig.\ 7).  Adopting a
dusty atmosphere with \teff\,=\,1900\,K, this limits coherent clear
clouds to a coverage of no more than 10--15\% and 200\,K cooler spots
to a 20\% coverage.
 
\begin{figure}[t]
\begin{center}
\epsfig{file=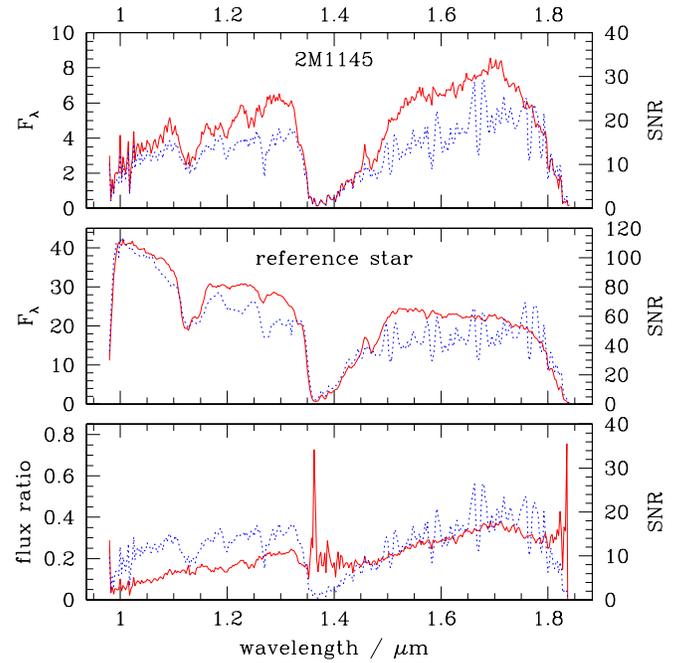, width=9cm, angle=0}
\end{center}
\caption{Near infrared spectra of the L1.5 dwarf 2M1145 (top), the reference
star observed in the same slit (middle) and their relative spectra
(bottom) used to monitor for variability independent of changes in the Earth's
atmosphere.  The red (solid) lines show the flux (left scale) and the
blue (dashed) lines the SNR (right scale). See Bailer-Jones (2002).}
\label{fig6}
\end{figure}

\begin{figure}[t]
\begin{center}
\epsfig{file=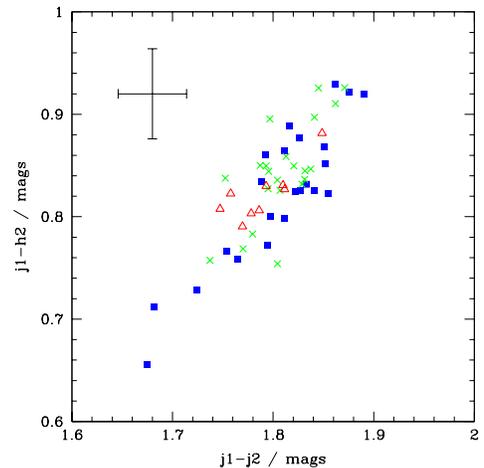, width=6.5cm, angle=0}
\end{center}
\caption{Colour correlated variability in 2M1145 (see Fig.\ 6). The
bands are defined in Fig.\ 5.}
\label{fig7}
\end{figure}

Ongoing work is aimed at better characterizing UCD variability and
achieving higher sensitivity.  Recent observations have been obtained
of several UCDs (see Fig.\ 8). I am also extending this work to
field M dwarfs and T dwarfs. Other useful observational signatures
include polarimetry and high resolution monitoring of line profiles
(Doppler imaging).

\begin{figure}[t]
\begin{center}
\epsfig{file=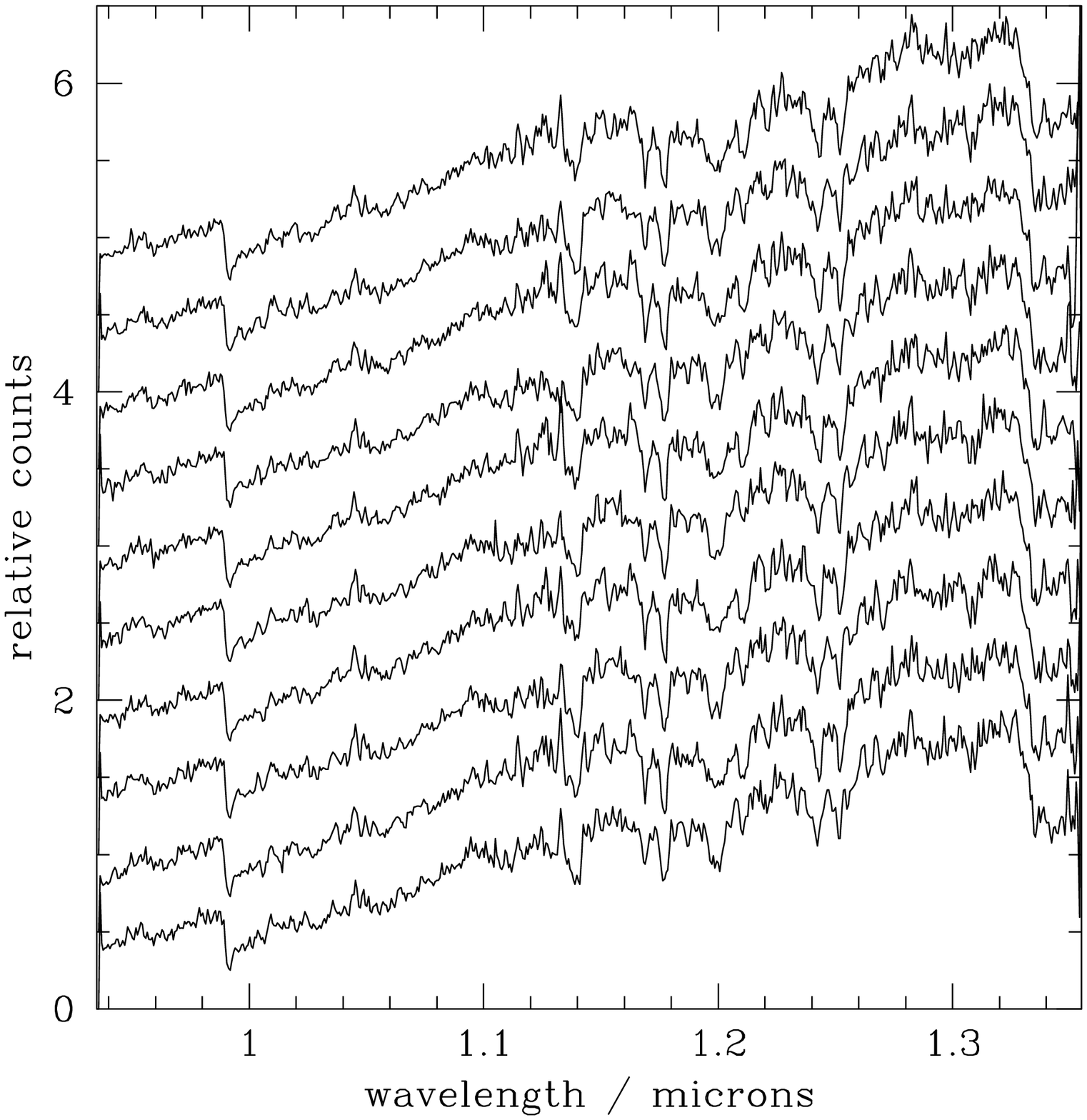, width=8cm, angle=0}
\end{center}
\caption{A sequence of ten relative spectra (from a total of 60) for
the L2V target SSSPM J0828-1309. Time increases
from bottom to top in steps of about four minutes and each relative
spectrum has been offset from the previous by 0.5.  One can see
variability at various places, e.g.\ around the Na{\small I} doublet
at 1.138\micron\ and 1.141\micron\ (which coincides with a water band)
and the K{\small I} doublet at 1.168\micron\ and 1.177\micron.}
\label{fig8}
\end{figure}

\end{document}